\documentclass[ ]{revtex4}
\usepackage{hyperref}
\usepackage{amsmath}

\usepackage{amsfonts}

\begin{document}
\title{\bf \Large   Quantum Gauge Freedom in Very Special Relativity}

 \author { Sudhaker Upadhyay }
 \email{sudhakerupadhyay@gmail.com}
\affiliation { Centre for Theoretical Studies,  
Indian Institute of Technology Kharagpur,  Kharagpur-721302, West Bengal, India}

 \author {Prasanta K. Panigrahi}
 \email{pprasanta@iiserkol.ac.in; panigrahi.iiser@gmail.com }
\affiliation {Indian Institute of Science Education and Research Kolkata, Mohanpur 741246, West Bengal, India}

\begin{abstract} 
We demonstrate Yokoyama gaugeon formalism  for the Abelian one-form gauge (Maxwell)  as well as for Abelian two-form gauge theory in the 
very special relativity (VSR) framework. In VSR scenario, the extended action due to introduction of gaugeon fields also
possesses form invariance under quantum gauge transformations.
It is observed that  the gaugeon field  together with gauge field 
 naturally acquire mass, which is different from the conventional Higgs mechanism. The quantum gauge transformation implements 
a shift in gauge parameter.  
Further, we analyse
the BRST symmetric gaugeon formalism in VSR which embeds only one  subsidiary
condition rather than two.
\end{abstract}
 
\maketitle
 \section{Overview and motivation}
In recent times,  the violations of Lorentz symmetry have  been studied with great interest
\cite{ko,ca,co,ca1,ar,co1}, though special relativity (SR), whose underlying  Lorentz symmetry  is valid at the largest energies available  these days \cite{pi}. 
However, the  violation of Lorentz symmetry
has been considered as a possible evidence for Planck
scale physics \cite{ja}.  In this context, Cohen and Glashow \cite{7} have  
proposed  that the laws of physics need
not be invariant under the full Lorentz group but rather under its
  subgroups that still preserves the basic elements of
SR, like the constancy of the velocity of light. Any scheme whose space-time symmetries
consist of translations along with any   Lorentz
subgroups is referred to as very special relativity (VSR).
Most common subgroups fulfilling the essential requirements are the  homothety group  $HOM(2)$ (with three parameters) and the similitude group $SIM(2)$ (with four parameters) \cite{7}. The generators of  $HOM(2)$  are $T_1=K_x+J_y$, $T_2=K_y -J_x$, and $K_z$, where $J$ and  $K$ are the generators of rotations and boosts, respectively. The generators of  $SIM(2)$ are $T_1=K_x+J_y$, $T_2=K_y -J_x$,  $K_z$  and $J_z$. These subgroups  will be enlarged to the full
Lorentz group  when supplemented with discrete space-time symmetries
$CP$. Recently, the three-dimensional supersymmetric Yang-Mills theory coupled to matter fields,
 (supersymmetric) Chern-Simons theory  in $SIM(1)$ superspace formalism \cite{voh}
 and $SIM(2)$ superspace formalism \cite{voh1} are derived.
The Feynman rules and supergraphs \cite{voh2}   in $SIM(2)$ superspace also has been studied.

  VSR admits natural origin to lepton-number conserving neutrino masses without the need for sterile
(right-handed) states \cite{8}. This implies that neutrinoless double beta decay is forbidden, if VSR is solely responsible for neutrino masses.
Further, VSR is generalized to
 $N = 1$ 
SUSY gauge theories  \cite{9}, where it is shown that these theories contain two conserved
supercharges rather than the usual four. VSR is also modified   by quantum corrections to produce a curved space-time with a cosmological constant \cite{10}, where it is shown that the symmetry group $ISIM(2)$ does admit a 2-parameter family of continuous
deformations, but none of these give rise to non-commutative translations analogous to those of the
de-Sitter deformation of the Poincar{\'e}  group. The   VSR is generalized to curved space-times also, where it has been found that 
gauging the $SIM(2)$ symmetry, which leaves the preferred
null direction   invariant, does not provide the complete couplings
to the gravitational background \cite{mu}. The three
subgroups relevant to VSR are also  realized in the non-commutative space-time \cite{jab,subi} and in this setting  the non-commutativity parameter $\theta^{\mu\nu}$ behaves as lightlike.
VSR has been generalized in various  contexts. For example,
 the generalization of VSR ideas to de Sitter
spacetime is studied where breaking of de Sitter invariance arises in two different
ways \cite{12}. This has also been shown that the event space underlying the
dark matter and the dark gauge fields supports the algebraic structure underlying VSR
\cite{13}. A generalization of VSR in cosmology is also proposed
where  an anisotropic modification to the Friedmann-Robertson-Walker (FRW) line element occurs
and for an arbitrarily oriented 1-form, the FRW space-time becomes of the Randers-Finsler type \cite{14}. The VSR modifications to  the quantum electrodynamics   and  the
massive spin-1 particle are reported in Refs. \cite{alv, 15}.
 Furthermore, the  generalization to the case of  non-Abelian gauge fields
 is made in \cite{16} and, in this context,
 the spontaneous symmetry-breaking mechanism to give a flavor-dependent VSR
mass to the gauge bosons is also studied.
 VSR is also studied as  background field theory, where 
 averaging observable    generates the nonlocal terms familiar from $SIM(2)$ theories, while the short-distance behavior of the background field fermion propagator generates the infinite number of higher-order vertices of $SIM(2)$-quantum electrodynamics \cite{ild}.
The electrostatic solutions  as well as the VSR dispersion relations
for Born--Infeld electrodynamics  are investigated to be of a massive particle with nonlinear
modifications in VSR scenario  \cite{buf}. 
Recently,  VSR generalization of the tensor field (reducible gauge) theories has
also been analyzed 
using a Batalin-Vilkovisy (BV) formulation   \cite{sud1}.
A rigorous construction of quantum field theory with a preferred direction is also studied 
very recently \cite{lee}. We would like to generalize the VSR in  gaugeon formalism 
as gaugeon formalism is important in studying quantum gauge symmetry as well as in renormalization
of gauge parameter. 

 The basic idea behind the gaugeon formalism \cite{yo0} is to introduce the so-called  gaugeon fields   to the action which represent quantum gauge freedom. Originally,  this formulation was developed in the case of  quantum electrodynamics to settle the issues of renormalization of gauge parameter. 
In this connection the occurrence of shift in gauge parameter
during renormalization  \cite{haya} was addressed naturally by  connecting theories in two different gauges within the
same family by a $q$-number gauge transformation \cite{yo0}. Further, this formalism has also been generalized to the case of Yang-Mills theory \cite{yo10}. It has been found that  gaugeon modes possess negative normed state  
which has been dealt with through Gupta-Bleuler type subsidiary condition. However, such condition is not applicable everywhere. For example, it fails when interaction is present between gaugeon fields.
To improve the situation further, such subsidiary condition has been replaced by Kugo-Ojima type restrictions with the help of BRST charge \cite{ki,mk,kugo, kugo1}.
The importance of BRST symmetry can be found in various contexts \cite{el0,el00,el,el1,el2}.
  Though gaugeon formalism  has been  studied for various theories 
\cite{ki, mk, mk1, naka, rko, miu, mir1, mir2,sud0,sud}, it is still unexplored in 
the  VSR context. We are interested to study the VSR effects on the gauge parameter, quantum gauge transformations and  on the Yokoyama subsidiary conditions.
This is the motivation of the present study.

Here, for illustration, we first consider Abelian 1-form (Maxwell) theory in VSR
($SIM(2)$-invariant) scenario and revisit 
the standard BRST quantization of the theory. Further to  discuss the quantum gauge freedom,
 we introduce the gaugeon fields to the VSR  action. Remarkably,
terms containing gaugeon fields in the action remain local, even after breaking the
full Lorentz invariance by some non-local terms. We found that
the resulting action in VSR framework does not respect form invariance
under the standard quantum gauge transformation. 
However, this action remains form-invariant under VSR-modified
quantum gauge transformations.  Also, the gauge parameter gets an automatic shift 
under this transformation even in VSR case. 
We also show that the gaugeon fields satisfy  
Proca equation, which signifies that these fields are massive. The introduction of 
gaugeon fields increases the physical degrees of freedom. To make it consistent with 
the original theory,  we impose the Gupta-Bleuler type subsidiary
condition, which removes the unphysical gaugeon modes from the theory. 
Further, to improve the situation with the Gupta-Bleuler type subsidiary,  which has certain 
limitations, we demonstrate the BRST symmetric gaugeon formalism 
by introducing ghosts corresponding to gaugeon fields in VSR, which yields the
more acceptable Kugo-Ojima subsidiary condition. This manifests the
validity of gaugeon formalism of 1-form gauge theory in VSR. 
The novel observation here is that unlike to SR invariant case, the gaugeon fields together with the gauge fields get mass  automatically. Furthermore, we explore
Sakoda's technique in VSR for changing the gauge parameter of the linear covariant
gauge from generating functional points of view with respect to the gauge freedom.
We then generalize the obtained results for the case of Abelian 2-form gauge theory. For this purpose, 
 we first consider Abelian 2-form gauge theory in VSR framework.
The Abelian 2-form gauge theory, in a similar fashion to 1-form case, is also
invariant under the modified gauge transformation. We   compute
the BRST symmetry of the theory.  Moreover, to get consistent description of quantum gauge freedom  for 2-form gauge theory, we introduce gaugeon vector fields to the 2-form action. The resulting
action remains invariant under VSR-modified quantum gauge symmetry. The gauge parameter 
translates under this symmetry for the 2-form case also,
which is consistent with the renormalized gauge parameter.
 We observe that the vector gaugeon fields become massive in 2-form case also.
We   study the BRST symmetric gaugeon formalism for Abelian 2-form gauge theory in VSR. In this context, we show that the two subsidiary conditions to
remove the unphysical fields are converted to a single but more general condition.
Thus, the present investigation ensures the validity of gaugeon formalism in VSR.

The paper is organized in the following manner. In Sec. II, we derive both the gaugeon and BRST 
symmetric gaugeon formalism  for Abelian 1-form gauge theory in VSR framework. In Sec. III, we 
generalize the discussion
for both the gaugeon and BRST symmetric gaugeon formalism for Abelian 2-form gauge 
theory  in VSR. We summarize our results   in the last section and indicate directions
for further study. 
 \section{1-form gauge theory in VSR}
We analyse below the partially BRST symmetric and fully BRST symmetric gaugeon formalisms  for 1-form gauge theory by extending the configuration space
 with the help of quantum fields. To make the theory consistent with the original 
 one we remove the redundant degrees of freedom due to gaugeon fields with the help
 of suitable subsidiary condition. 
  \subsection{Brief review}
  In this subsection, we recapitulate the $SIM(2)$-invariant 1-form gauge theory in VSR \cite{15,16}.
  The most general gauge invariant action, quadratic in the gauge field, is given by
\begin{equation} 
  S = \int d^d x \left( - \frac{1}{4} \tilde{F}_{\mu \nu} \tilde{F}^{\mu \nu}
  + \frac{g}{2} \frac{1}{n\cdot  \partial} n_{\alpha} \tilde{F}^{\alpha \mu}
  \frac{1}{n\cdot  \partial} n_{\nu} \tilde{F}^{\nu}_{\mu}  \right),  \label{Abelianaction}
\end{equation}
where  $n_{\alpha}$ is a chosen preferred null direction that transforms
multiplicatively under a VSR transformation, $g$ is a constant, and $\tilde{F}_{\mu \nu}$ is wiggle field-strength tensor defined as
  \begin{equation}
 \tilde{F}_{\mu \nu} =\partial_{\mu} A_{\nu}-\partial_{\nu} A_{\mu} - \frac{1}{2} \frac{m^2}{n\cdot
  \partial} n_{\mu} A_{\nu}+\frac{1}{2} \frac{m^2}{n\cdot
  \partial} n_{\nu} A_{\mu}.
\end{equation}
Here, the VSR mass parameter $m$ sets the scale for the VSR effects and was introduced by 
a dimensional reason. The above action is not invariant under the standard gauge transformation. Rather, it is invariant under a VSR type gauge transformation: 
\begin{equation} \label{1}
  \delta A_{\mu} = \partial_{\mu} \Lambda - \frac{1}{2} \frac{m^2}{n\cdot
  \partial} n_{\mu} \Lambda,
\end{equation}
where $\Lambda$ is a local transformation parameter.

We handle the kind of  non-local terms,  present above,  with the help of following definition \cite{ale}:
\begin{eqnarray}
\frac{1}{n\cdot\partial} =\frac{1}{\partial_t +\partial_z} = \int_0^\infty da\ e^{-an\cdot\partial}.
\end{eqnarray}

 From action (\ref{Abelianaction}) (with $g=0$), the equations of motion (EOM) for   free Abelian field  
  is calculated by,   
\begin{equation}
\partial^{\mu}  \tilde F_{\mu \nu}- \frac{1}{2} \frac{m^2}{n\cdot
  \partial} n^{\mu} \tilde F_{\mu \nu} = 0,\label{eqn}
\end{equation}
which further leads to
   \begin{eqnarray*}
    ( \Box - m^2) A_{\nu} = 0, 
\end{eqnarray*}
for a VSR type Lorentz gauge condition,
$
\partial^{\mu} A_{\mu}- \frac{1}{2} \frac{m^2}{n\cdot
  \partial} n^{\mu}  A_{\mu} = 0
$. 
This implies that the gauge field $A_\mu$ has a mass $m$ and the action (\ref{Abelianaction}) describes a massive gauge field.

To  achieve  the VSR-type Lorentz gauge in quantum action, we need to add  the following 
term
to the action (\ref{Abelianaction}):
\begin{eqnarray*}
  S^L_{{gf} + {gh}} =    \int d^d x \left\{  B \left(
   {\partial}_{\mu} A^{\mu}- \frac{1}{2} \frac{m^2}{n\cdot
  \partial} n_{\mu}A^{\mu} + \frac{1}{2} \alpha B \right)  + i c_\star (\Box - m^2)c   \right\},
  \label{gfs}
\end{eqnarray*}
where the last term is induced ghost terms with   mass $m$. 
 We note here that the gauge-fixed action in VSR-type axial gauge can also be 
 constructed which has rather simpler form than the Lorenz gauge \cite{sud1}.

The  effective action in VSR-type Lorentz gauge is given by,
 \begin{equation}
 S_{eff}= S+  S^L_{{gf} + {gh}}, \label{effe}
\end{equation} which is not invariant under (VSR-type) gauge transformation 
but remains invariant under the following 
 BRST transformation:
 \begin{eqnarray}
s_b A_{\mu} &=&   \partial_{\mu} c - \frac{1}{2} \frac{m^2}{n\cdot
  \partial} n_{\mu} c,  \nonumber \\
s_b c &=& 0,  \nonumber\\
 s_b  c_\star &=&  i B,   \nonumber\\
s_b B &=& 0.
\end{eqnarray}
The gauge-fixed action  is BRST exact as it is evident from,
\begin{eqnarray*}
  S^L_{{gf} + {gh}}  =  - i s_b \int d^d x  \left\{  c_\star \left(
  {\partial}_{\mu} A^{\mu}- \frac{1}{2} \frac{m^2}{n\cdot
  \partial} n_{\mu}A^{\mu}  + \frac{1}{2} \alpha B \right)  
  \right\}.
\end{eqnarray*}
We note that  this gauge-fixed action,  added to modify the classical action,
 has no contribution to the physical matrix elements of the theory.  All the physical matrix elements of the theory are independent of the choice of the gauge-fixing parameter
$\alpha$.

Next, we discuss the gaugeon formalism for 1-form gauge theory in VSR.
\subsection{Gaugeon formalism}
Following the standard technique, we introduce  the gaugeon field $Y$ and its associated field $Y_\star$, obeying Bose-Einstein statistics, to the effective action (\ref{effe}). With such introduction, the Yokoyama effective action for the Abelian one-form gauge theory in VSR follows,
\begin{eqnarray}
  S_Y &=& \int d^d x \left( - \frac{1}{4} \tilde{F}_{\mu \nu} \tilde{F}^{\mu \nu}
  + \frac{g}{2} \frac{1}{n\cdot  \partial} n_{\alpha} \tilde{F}^{\alpha \mu}
  \frac{1}{n\cdot  \partial} n_{\nu} \tilde{F}^{\nu}_{\mu} +
  \partial_\mu BA^\mu -\frac{1}{2}\frac{m^2}{n\cdot\partial}n_\mu B A^\mu \right.\nonumber\\
  &-&\left.   Y_\star [\square-m^2] Y  +\frac{\varepsilon}{2} (Y_\star +\alpha B)^2+i c_\star  [\square-m^2] c   \right),   \label{ga}
\end{eqnarray}   
 where $\alpha$ is the  gauge-fixing parameter and $\varepsilon (\pm)$
is the sign factor. Here, we note that VSR effect does not change the   spin-statistics theorem for the fields.
The gaugeon action  (\ref{ga}) remains form invariant under following   quantum gauge transformations:
 \begin{eqnarray}
&&  \delta_q {A}_\mu = \tau\left(\partial_\mu Y-\frac{1}{2}\frac{m^2}{n\cdot\partial}n_\mu Y\right),
 \nonumber\\
 && \delta_q {Y}_\star = -\tau B,\nonumber\\
 &&\delta_q {B}=0,\ \ \delta_q {Y}=0,\nonumber\\
 &&\delta_q {c}=0,\  \ \ \delta_q {c}_\star =0,
 \end{eqnarray}
 with a shift in gauge parameter
 \begin{equation}
 \hat{\alpha}=\alpha+\tau.
 \end{equation}
The two subsidiary conditions to remove the unphysical modes are
 \begin{eqnarray}
 Q|\mbox{phys}\rangle =0,\nonumber\\
 Y_\star^{(+)}|\mbox{phys}\rangle =0.
 \end{eqnarray}
 The first condition is the usual one of the standard formalism, which confines
the unphysical gauge  modes by the quartet mechanism. The second condition removes the gaugeon
modes. 
 The decomposition in positive and negative frequency parts are valid because of the free equation:
 \begin{equation}
(\square -m^2)Y_\star =0.\label{ma}
 \end{equation} 
 If this does not hold, the positive frequency part  becomes ambiguous and the second subsidiary condition   contradicts with time evolution.
  Eq. (\ref{ma})  reflects that the gaugeon field   gets mass along with 
 gauge and ghost fields in VSR.
 The photon two-point function  in VSR is calculated by
\begin{eqnarray}
D_{\mu\nu} =\frac{1}{(k^2 +m^2)^2}\left[(k^2 +m^2)g_{\mu\nu} +(\epsilon\alpha^2 -1)\left(
k_\mu k_\nu +\frac{1}{2}m^2 \left(\frac{k_\mu n_\nu +k_\nu n_\mu}{n\cdot k}\right) 
+\frac{1}{4} m^4\frac{n_\mu n_\nu}{(n\cdot k)^2}\right) \right],
\end{eqnarray}
where $g=0$ is considered for simplicity. It can be seen that the ultra-violet behavior of
 above two-point function $\left\langle A_\mu A_\nu \right\rangle$
is $O(1/k^2)$, which coincides the standard two-point function of SR  invariant theory.
 \subsection{BRST symmetric gaugeon formalism }
 In this subsection, we discuss the BRST symmetric gaugeon formalism in VSR. For this purpose, we first extend the effective Yokoyama
Lagrangian density (\ref{ga}), by introducing two
Faddeev-Popov ghosts $K, K_\star$ corresponding to the gaugeon fields as follows:
 \begin{eqnarray}
  S_{BY} &=& \int d^d x \left( - \frac{1}{4} \tilde{F}_{\mu \nu} \tilde{F}^{\mu \nu}
  + \frac{g}{2} \frac{1}{n\cdot  \partial} n_{\alpha} \tilde{F}^{\alpha \mu}
  \frac{1}{n\cdot  \partial} n_{\nu} \tilde{F}^{\nu}_{\mu} +
  \partial_\mu BA^\mu -\frac{1}{2}\frac{m^2}{n\cdot\partial}n_\mu B A^\mu \right.\nonumber\\
  &-&\left.   Y_\star [\square-m^2] Y  +\frac{\varepsilon}{2} (Y_\star +\alpha B)^2+i c_\star  [\square-m^2] c  +i K_\star  [\square-m^2] K \right).  
\end{eqnarray}  
The BRST symmetry for the above action is written by,
\begin{eqnarray}
s_b A_\mu &=& \partial_\mu c - \frac{m^2}{n\cdot  \partial} n_{\mu}c,\nonumber\\
s_b c_\star &=& -iB,\ \
s_b c  =0,\nonumber\\
s_b B  &=&0,\ \
s_b Y  =K,\nonumber\\
s_b K_\star  &=& -iY_\star,\ \
s_b K  =0,\nonumber\\
s_b Y_\star  &=&0. \label{brs0}
\end{eqnarray}
The BRST charge using Noether's theorem is calculated by
\begin{eqnarray}
Q_B =\int d^{d-1}x (c\overleftrightarrow{\tilde\partial_0} B +K\overleftrightarrow{\tilde\partial_0} Y_\star),\label{bch}
\end{eqnarray}
where $\tilde\partial_0 =\partial_0 - \frac{m^2}{n\cdot  \partial} n_{0}$.

With the help of this BRST charge, we can define the physical subspace of the total Hilbert space satisfying,
\begin{eqnarray}
Q_B|\mbox{phys}\rangle =0.
\end{eqnarray}
It is interesting to observe that,   even after getting extended by introduction of ghosts corresponding to
gaugeon fields, the action
  admits quantum gauge transformation given by,
\begin{eqnarray}
\delta_q A_\mu &=& \tau \left(\partial_\mu Y - \frac{m^2}{n\cdot  \partial} n_\mu Y\right),\nonumber\\
\delta_q Y_\star &=& -\tau B,\ \
\delta_q B = 0,\nonumber\\
\delta_q Y &=& 0,\ \
\delta_q c = \tau K,\nonumber\\
\delta_q  K_\star &=& -\tau c_\star,\ \
\delta_q c_\star = 0,\ \
\delta_q K = 0.\label{qq}
\end{eqnarray}
These transformations only shift   gauge parameter leaving the action form-invariant.
We observe that the above quantum gauge transformation   commutes with the
BRST transformation. This ensures that the BRST charge (\ref{bch}) is invariant under the
quantum gauge transformation:
\begin{equation}
\delta_q Q_B =0.
\end{equation}
 Therefore, the physical subspace  is  invariant under the quantum gauge transformation,
 as the physical subspace can be constructed with the help of BRST charge.
However, the BRST symmetric gaugeon formalism is more acceptable in the sense that,  this situation does not occur in  partially BRST symmetric theory described in subsection B.
\subsection{Sakoda's extension of the gauge freedom of the vector field in VSR}
In this subsection, as a first step, we consider the Maxwell gauge field   and starting with the
 generating functional for 1-from  in the Landau gauge, we extend the gauge freedom by   applying
the Harada-Tsutsui gauge recovery procedure in the VSR context.
Let us start with the effective action for the Maxwell theory in VSR-type Landau gauge 
(Eq. (\ref{effe}) with $\alpha=0$) given by
\begin{eqnarray}
S_{eff} = \int d^d x \left[ - \frac{1}{4} \tilde{F}_{\mu \nu} \tilde{F}^{\mu \nu}
  + \frac{g}{2} \frac{1}{n\cdot  \partial} n_{\nu} \tilde{F}^{\nu \mu}
  \frac{1}{n\cdot  \partial} n_{\alpha} \tilde{F}^{\alpha}_{\mu} +B \left(
   {\partial}_{\mu} A^{\mu}- \frac{1}{2} \frac{m^2}{n\cdot
  \partial} n_{\mu}A^{\mu}  \right)  + i c_\star (\Box - m^2)c  \right].
\end{eqnarray}
Now, the generating functional corresponding to this action
 is expressed by
\begin{eqnarray}
Z=\int {\cal D}A{\cal D}B{\cal D}c_\star Dc\ e^{iS_{eff}} =\int {\cal D}A{\cal D}B\ {\cal W}_0,
\end{eqnarray}
with 
\begin{eqnarray}
{\cal W}_0 =\Delta\  e^{i \int d^d x \left[ - \frac{1}{4} \tilde{F}_{\mu \nu} \tilde{F}^{\mu \nu}
  + \frac{g}{2} \frac{1}{n\cdot  \partial} n_{\alpha} \tilde{F}^{\alpha \mu}
  \frac{1}{n\cdot  \partial} n_{\nu} \tilde{F}^{\nu}_{\mu} +B \left(
   {\partial}_{\mu} A^{\mu}- \frac{1}{2} \frac{m^2}{n\cdot
  \partial} n_{\mu}A^{\mu}  \right) \right]},\label{gh}
\end{eqnarray}
where $\Delta =\mbox{det}(\Box - m^2)$.
Since the action in (\ref{gh}) contains the gauge fixing part   together with the classical
part, therefore,  the functional ${\cal W}_0$ is not   invariant under the
VSR-type gauge transformation,
\begin{eqnarray}  
  \delta A_{\mu} &=& \partial_{\mu} \Lambda - \frac{1}{2} \frac{m^2}{n\cdot
  \partial} n_{\mu} \Lambda,\nonumber\\
   \delta B&=&0.
\end{eqnarray}
In order to study the VSR-type gauge invariance of this gauge-fixed functional (quantum action), we promote the function $\Lambda(x)$  to a dynamical variable and
define
\begin{eqnarray}
{\cal W}'_0 =\Delta\  e^{i \int d^d x \left[ - \frac{1}{4} \tilde{F}_{\mu \nu} \tilde{F}^{\mu \nu}
  + \frac{g}{2} \frac{1}{n\cdot  \partial} n_{\alpha} \tilde{F}^{\alpha \mu}
  \frac{1}{n\cdot  \partial} n_{\nu} \tilde{F}^{\nu}_{\mu} +B \left(
   {\partial}_{\mu} A^{\mu}- \frac{1}{2} \frac{m^2}{n\cdot
  \partial} n_{\mu}A^{\mu}  +(\Box - m^2)\Lambda\right) \right]}.
\end{eqnarray}
This functional ${\cal W}'_0$  is found invariant under the extended VSR-type gauge transformation,
\begin{eqnarray}  
  \delta A_{\mu} &=& \partial_{\mu} \theta - \frac{1}{2} \frac{m^2}{n\cdot
  \partial} n_{\mu} \theta,\nonumber\\
    \delta \Lambda &=& -\theta,\nonumber\\
   \delta B&=&0.
\end{eqnarray}
Because of this (quantum) gauge symmetry, the generating functional for ${\cal W}'_0$,
defined as
\begin{eqnarray}
Z=\int {\cal D}A{\cal D}B{\cal D}\Lambda \  {\cal W}'_0,
\end{eqnarray}
is divergent. In order to remove the extra degree of freedom, 
we need to fix the VSR-type gauge for $\Lambda$. Here, we consider the following VSR-type 
gauge-fixing condition:
\begin{eqnarray}
-(\Box - m^2)\Lambda =C.
\end{eqnarray}
Utilizing Faddeev-Popov trick, we write the generating functional as
\begin{eqnarray}
Z=\int {\cal D}A{\cal D}B{\cal D}\Lambda \  {\cal W}'_0 \Delta \delta (-\Box\Lambda +m^2\Lambda -C),
\end{eqnarray}
as $\Delta \delta (-\Box\Lambda +m^2\Lambda -C) =1$.
By writing the   Fourier integral for the delta functional   with respect to $B_\Lambda$
and applying 't Hooft averaging
with a Gaussian weight,  the generating functional reduces to,
\begin{eqnarray}
Z=\int {\cal D}A{\cal D}B{\cal D}\Lambda {\cal D}B_\Lambda \  {\cal W}_1,
\end{eqnarray}
where
\begin{eqnarray}
{\cal W}_1 =\Delta\Delta\  e^{i \int d^d x \left[ - \frac{1}{4} \tilde{F}_{\mu \nu} \tilde{F}^{\mu \nu} + \frac{g}{2} \frac{1}{n\cdot  \partial} n_{\alpha} \tilde{F}^{\alpha \mu}
  \frac{1}{n\cdot  \partial} n_{\nu} \tilde{F}^{\nu}_{\mu}
  +B \left(
   {\partial}_{\mu} A^{\mu}- \frac{1}{2} \frac{m^2}{n\cdot
  \partial} n_{\mu}A^{\mu}  +(\Box - m^2)\Lambda\right) -B_\Lambda (\Box - m^2) \Lambda +
  \frac{a}{2}B_\Lambda^2\right]}.
\end{eqnarray}
The two Faddeev-Popov determinants in VSR are calculated as
\begin{eqnarray}
\Delta\Delta=\int {\cal D}c_\star {\cal D}c {\cal D}\eta_\star {\cal D}\eta \ e^{i \int d^d x \left[ic_\star(\Box - m^2)c +i\eta_\star(\Box - m^2)\eta \right]}.
\end{eqnarray}
Consequently, we obtain the following Sakoda effective action for the   extension of the gauge freedom:
\begin{eqnarray}
S_{Sakoda} &=& \int d^d x \left[ - \frac{1}{4} \tilde{F}_{\mu \nu} \tilde{F}^{\mu \nu} + \frac{g}{2} \frac{1}{n\cdot  \partial} n_{\alpha} \tilde{F}^{\alpha \mu}
  \frac{1}{n\cdot  \partial} n_{\nu} \tilde{F}^{\nu}_{\mu}
  +B \left(
   {\partial}_{\mu} A^{\mu}- \frac{1}{2} \frac{m^2}{n\cdot
  \partial} n_{\mu}A^{\mu}  +(\Box - m^2)\Lambda\right)\right.\nonumber\\
  & -&\left. B_\Lambda (\Box - m^2) \Lambda +
  \frac{a}{2}B_\Lambda^2 +ic_\star(\Box - m^2)c +i\eta_\star(\Box - m^2)\eta \right].\label{sk}
\end{eqnarray}
This action enjoys the following extended VSR-type BRST transformation:
\begin{eqnarray}
s_b A_{\mu} &=&   \partial_{\mu} c - \frac{1}{2} \frac{m^2}{n\cdot
  \partial} n_{\mu} c, \ \ s_b \Lambda =-(c+\eta ), \nonumber \\
s_b c &=& 0, \ \ s_b\eta =0, \ \  
 s_b  c_\star = i B_\Lambda,   \nonumber\\
s_b \eta_\star &=& i(B_\Lambda -B), \ \ s_bB_\Lambda =0, \ \ s_bB =0.
\end{eqnarray}
  Sakoda states that the effective  action for the   extension of the gauge freedom must
 be equivalent to the action of the gaugeon formalism \cite{sako}.
Remarkably, the  Sakoda effective action  (\ref{sk}) coincides with
the gaugeon action (\ref{ga}) in VSR framework after redefining the fields as, 
\begin{eqnarray}
\Lambda =\alpha Y,\ \ B_\Lambda =\frac{1}{\alpha}Y_\star +B,\ \ \eta= K,\ \
\eta_\star =K_\star.
\end{eqnarray}
Here $\alpha$ is a numerical parameter satisfying $a=\epsilon\alpha^2$.
Thus, we note  that the field $\Lambda(x)$ introduced as an extended gauge freedom
plays the character of a gaugeon field in VSR.
\section{Abelian 2-form gauge theory in VSR}
We start with  
  the field-strength tensor in VSR for Kalb-Ramond tensor field $B_{\mu\nu}$ \cite{kal} in VSR involving a fixed null vector $n_\mu$:  
 \begin{eqnarray}
 F_{\mu\nu\rho}&=&\partial_\mu B_{\nu\rho}+\partial_\nu B_{\rho\mu}+\partial_\rho B_{\mu\nu}
+\frac{1}{2}m^2\left[n_\mu \frac{1}{(n\cdot\partial)^2} n^\alpha(\partial_\nu B_{\rho\alpha} +\partial_\rho B_{\nu\alpha}) \right.\nonumber\\
 &+&\left. n_\nu \frac{1}{(n\cdot\partial)^2} n^\alpha(\partial_\rho B_{\mu\alpha} +\partial_\mu B_{\rho\alpha})+n_\rho \frac{1}{(n\cdot\partial)^2} n^\alpha(\partial_\mu B_{\nu\alpha} +\partial_\nu B_{\mu\alpha})\right].\label{fi}
 \end{eqnarray}
 As before, the null vector  $n^\mu$  transforms multiplicatively under a VSR transformation
to ensure the invariance of  non-local terms.
 This field-strength tensor  is not invariant under the standard gauge transformation 
 $ \delta B_{\mu\nu}= \partial_\mu \zeta_\nu - \partial_\nu\zeta_\mu$, where  $\zeta_{\mu}(x)$ is a vector  
parameter. One can  dualize a two-form to a pseudoscalar in the VSR also satisfying 
modified Bianchi identity.
Rather, it remains invariant under the following modified (VSR-type) gauge transformation: 
 \begin{eqnarray}
 \delta B_{\mu\nu}&=&\tilde\partial_\mu \zeta_\nu -\tilde\partial_\nu\zeta_\mu,\nonumber\\
 &=&\partial_\mu\zeta_\nu -\partial_\nu\zeta_\mu -\frac{1}{2}\frac{m^2}{n\cdot\partial}n_\mu\zeta_\nu +
 \frac{1}{2}\frac{m^2}{n\cdot\partial}n_\nu\zeta_\mu.
 \end{eqnarray}
As before, to have the usual mass dimension for  wiggle operator, $
  \tilde{\partial}_{\mu} = \partial_{\mu} - \frac{1}{2} \frac{m^2}{n\cdot
  \partial} n_{\mu},
$ a constant $m$ has to be introduced.

The gauge invariant action in VSR to describe the massive Kalb-Ramond tensor field is given by
 \begin{eqnarray}
 S_0=\frac{1}{12}\int d^dx\ \tilde F_{\mu\nu\rho}\tilde F^{\mu\nu\rho},\label{cl}
 \end{eqnarray}
 where the wiggle field-strength tensor has the following form:
 \begin{eqnarray}
 \tilde F_{\mu\nu\rho}&=&\tilde \partial_\mu B_{\nu\rho}+\tilde\partial_\nu B_{\rho\mu}+\tilde\partial_\rho B_{\mu\nu},\nonumber\\
 &=&\partial_\mu B_{\nu\rho}+\partial_\nu B_{\rho\mu}+\partial_\rho B_{\mu\nu}
 -\frac{1}{2}\frac{m^2}{n\cdot \partial}n_\mu B_{\nu\rho}-\frac{1}{2}\frac{m^2}{n\cdot \partial}n_\nu B_{\rho\mu}-\frac{1}{2}\frac{m^2}{n\cdot \partial}n_\rho B_{\mu\nu},\nonumber\\
 &=&F_{\mu\nu\rho}  -\frac{1}{2} m^2 \left(n_\mu\frac{1}{(n\cdot\partial)^2}n^\alpha F_{\nu\rho\alpha}+n_\nu\frac{1}{(n\cdot\partial)^2}n^\alpha F_{ \rho\mu\alpha}+n_\rho\frac{1}{(n\cdot\partial)^2}n^\alpha F_{\mu\nu\alpha} \right).
 \end{eqnarray}
It is evident from the above relation that, $\tilde F_{\mu\nu\rho}$ does not coincide with  $  F_{\mu\nu\rho}$ given in (\ref{fi}).
 
 The  EOM  for Kalb-Ramond field is calculated as,
 \begin{eqnarray}
 \tilde\partial_{\mu}\tilde F^{\mu\nu\rho}=0.
 \end{eqnarray}
 For the VSR-type Lorentz gauge $\tilde \partial_\mu B^{\mu\nu}=0$, the EOM reduces to
 \begin{eqnarray}
 [\square -m^2 ]B^{\nu\rho}=0,
 \end{eqnarray}
which remarkably implies that the field $B_{\mu\nu}$ has mass $m$. 

The   VSR-type Lorentz gauge can be implemented in the classical action by adding  suitable
gauge fixing and ghost terms.
The gauge fixing and ghost action for antisymmetric rank 2 tensor field in VSR-type Lorentz gauge is given by
\begin{eqnarray}
S_{gf+gh} &=&\int d^dx\left[ i\bar\rho_\nu \tilde\partial_\mu(\tilde\partial^\mu\rho^\nu -
\tilde\partial^\nu\rho^\mu )-\bar\sigma\tilde\partial_\mu\tilde\partial^\mu\sigma +\beta_\nu(\tilde\partial_\mu B^{
\mu\nu} +\lambda_1\beta^\nu -\tilde\partial^\nu\varphi)\right.\nonumber\\ 
&-&\left. i\bar\chi(\tilde\partial_\mu\rho^\mu +\lambda_2 \chi) -i\bar\rho^\mu \tilde \partial_\mu \chi \right],\nonumber\\
&=&\int d^dx\left[i\bar\rho_\nu \left(\partial_\mu\partial^\mu \rho^\nu -\partial_\mu\partial^\nu
\rho^\mu -m^2\rho^\nu +\frac{1}{2}	\frac{m^2}{n\cdot \partial}n^\nu\partial\cdot\rho
+ \frac{1}{2}	\frac{m^2}{n\cdot \partial} \partial^\nu n\cdot\rho\right.\right.\nonumber\\
& -&\left.\left. \frac{1}{4}\frac{m^4}{(n\cdot\partial)^2}n^\nu n\cdot\rho\right) -\bar{\sigma}
(\partial_\mu\partial^\mu -m^2)\sigma +\beta_\nu\partial_\mu B^{
\mu\nu} -\frac{1}{2}m^2\beta_\nu\frac{1}{n\cdot\partial}n_\mu B^{\mu\nu}+\lambda_1\beta_\nu\beta^\nu \right.\nonumber\\
&-& \left. \beta_\nu\partial^\nu\varphi +  \frac{1}{2}	\frac{m^2}{n\cdot \partial} \beta_\nu
  n^\nu\varphi  -i\bar\chi \partial_\mu\rho^\mu +\frac{i}{2}m^2\bar\chi\frac{1}{n\cdot\partial}n_\mu \rho^\mu-i\lambda_2\bar\chi\chi -i\bar\rho^\mu\partial_\mu\chi
 \right.\nonumber\\
&-& \left.\frac{i}{2}\frac{m^2}{n\cdot\partial}\bar\rho^\mu n_\mu\chi\right], \label{gfix}
\end{eqnarray}
where $\lambda_1$ and $\lambda_2$ are gauge parameters.
It is evident from the above expression that the ghost fields and ghost of ghost fields
have   mass $m$ in  VSR as well.
\subsection{Gaugeon formalism }
In this subsection, we  study the
Yokoyama gaugeon formalism to analyse the quantum gauge freedom for the Abelian rank-2 tensor field theory. 
We start with the effective Lagrangian density for a $d$-dimensional  theory in Landau
 gauge,
\begin{eqnarray}
S_{Y} &=&\int d^dx \left[\frac{1}{12}  \tilde F_{\mu \nu \rho}\tilde F^{\mu \nu \rho}  -i \tilde\partial_\mu\bar\rho_\nu ( \tilde\partial^\mu\rho^\nu - \tilde
\partial^\nu\rho^\mu )+ \tilde\partial_\mu\bar\sigma \tilde\partial^\mu\sigma +\beta_\nu( \tilde\partial_\mu B^{
\mu\nu}   - \tilde\partial^\nu\varphi)+\epsilon(Y^\star_\nu +\alpha\beta_\nu)^2\right. \nonumber\\ 
&-&\left. ( \tilde\partial_\mu Y^\star_\nu -
 \tilde\partial_\nu Y^\star_\mu ) \tilde\partial^\mu Y^\nu - i\bar\chi \tilde\partial_\mu\rho^\mu -i\chi ( \tilde\partial_\mu\bar\rho^\mu -
\lambda_2\bar\chi) \right], \nonumber\\
&=&\int d^dx \left[\frac{1}{12}  \tilde F_{\mu \nu \rho}\tilde F^{\mu \nu \rho}    -
i\bar\rho_\nu \left(\partial_\mu\partial^\mu \rho^\nu -\partial_\mu\partial^\nu
\rho^\mu -m^2\rho^\nu +\frac{1}{2}	\frac{m^2}{n\cdot \partial}n^\nu\partial\cdot\rho
+ \frac{1}{2}	\frac{m^2}{n\cdot \partial} \partial^\nu n\cdot\rho\right.\right.\nonumber\\
& -&\left.\left. \frac{1}{4}\frac{m^4}{(n\cdot\partial)^2}n^\nu n\cdot\rho\right) -\bar{\sigma}
(\partial_\mu\partial^\mu -m^2)\sigma +\beta_\nu\partial_\mu B^{
\mu\nu} -\frac{1}{2}m^2\beta_\nu\frac{1}{n\cdot\partial}n_\mu B^{\mu\nu}+\epsilon(Y^\star_\nu +\alpha\beta_\nu)^2 \right.\nonumber\\
&-& \left. \beta_\nu\partial^\nu\varphi  +  \frac{1}{2}	\frac{m^2}{n\cdot \partial} \beta_\nu
  n^\nu\varphi  -  (\partial_\mu Y^\star_\nu - \partial_\nu Y^\star_\mu )   \partial^\mu Y^\nu +\frac{1}{2}  (\partial_\mu Y^\star_\nu - \partial_\nu Y^\star_\mu )\frac{m^2}{n\cdot \partial} n^\mu Y^\nu  \right.\nonumber\\
&+& \left. \frac{1}{2}\frac{m^2}{n\cdot \partial} (n_\mu Y_\nu^\star -n_\nu Y_\mu^\star) \partial^\mu Y^\nu   -\frac{1}{4}\frac{m^2}{n\cdot \partial} (n_\mu Y_\nu^\star -n_\nu Y_\mu^\star)  \frac{m^2}{n\cdot \partial} n^\mu Y^\nu  -i\bar\chi \partial_\mu\rho^\mu +\frac{i}{2}m^2\bar\chi\frac{1}{n\cdot\partial}n_\mu \rho^\mu\right.\nonumber\\
&-& \left.  i\lambda_2\bar\chi\chi -i\bar\rho^\mu\partial_\mu\chi-\frac{i}{2}\frac{m^2}{n\cdot\partial}\bar\rho^\mu n_\mu\chi\right],
\label{ym}
\end{eqnarray}
where $Y_\nu$ and $Y_\nu^\star$
are the gaugeon fields respectively.

The Lagrangian density (\ref{ym}) is invariant under the following  BRST transformations: 
\begin{eqnarray}
s_b B_{\mu\nu} &=& ( \partial_\mu\rho_\nu -  \partial_\nu\rho_\mu -\frac{1}{2}	\frac{m^2}{n\cdot \partial}n_\mu \rho_\nu +\frac{1}{2}	\frac{m^2}{n\cdot \partial}n_\nu \rho_\mu),  \nonumber\\
s_b\rho_\mu &=& -i \partial_\mu\sigma +i\frac{1}{2}	\frac{m^2}{n\cdot \partial}n_\mu \sigma,   \   \ \ \ \ s_b\sigma 
= 0, \nonumber\\
s_b\bar\rho_\mu &=&i\beta_\mu ,   \ \ \    
s_b\beta_\mu = 0,\ \ \  
s_b\bar\sigma = \bar\chi ,   \ \ \ 
s_b\bar\chi =0,\nonumber\\
s_b\varphi &=&  \chi ,  \ \ \ \ s_b\chi =0, \ \ \ 
 s_b Y = 0,  \ \  \ s_b Y_\star =0.\label{brst11}
\end{eqnarray}
Now, we demonstrate the following quantum gauge transformation, under which  the Lagrangian density  (\ref{ym}) remains form-invariant:
\begin{eqnarray}
 &&\delta_q  B_{\mu\nu}  = \tau\left(  \partial_\mu Y_\nu -  \partial_\nu Y_\mu -\frac{1}{2}\frac{m^2}{n\cdot\partial}n_\mu Y_\nu +\frac{1}{2}\frac{m^2}{n\cdot\partial}n_\nu Y_\mu\right),  \nonumber\\
 &&\delta_q  Y_\nu^\star = -\tau \beta_\nu,\nonumber\\
 &&\delta_q  \Omega =0, \ \ \ \ \Omega =  \rho_\mu,  \sigma, \bar\rho_\mu, \beta_\mu, \bar\sigma,  \bar\chi, \varphi, \chi,  Y_\nu,\label{qq1}
 \end{eqnarray}
where $\tau$ is an infinitesimal transformation parameter.
The form-invariance of the Lagrangian density (\ref{ym}) under the quantum gauge transformation (\ref{qq1}) 
reflects the following  shift  in parameter:
\begin{equation}
\alpha \longrightarrow \hat\alpha  =\alpha +\tau \alpha.\label{alp}
\end{equation}
Further, according to Yokoyama, to remove the unphysical gauge and gaugeon  modes of the theory and to define physical states, one  imposes  
two  subsidiary conditions (the Kugo-Ojima type and Gupta-Bleuler type) as,
 \begin{eqnarray}
 Q_B|\mbox{phys}\rangle &=&0,\nonumber\\
 (Y_\nu^\star)^{(+)}|\mbox{phys}\rangle &=&0,\label{con}
 \end{eqnarray}
where $Q_B$ is the BRST charge. The expression for BRST charge using Noether's theorem is given by
\begin{eqnarray}
Q_B =\int d^{d-1}x \left[-2 \tilde F^{ 0\nu \rho} ( \tilde\partial_0\rho_\nu -  \tilde\partial_\nu\rho_0) + \beta_\nu ( \tilde\partial^0\rho^\nu -  \tilde\partial^\nu\rho^0) -  \tilde\partial_\nu \sigma ( \tilde\partial^0\bar\rho^\nu -  \tilde\partial^\nu\bar\rho^0) + \bar\chi  \tilde\partial^0\sigma - \chi B^0 \right].
\end{eqnarray}
 The   Kugo-Ojima type subsidiary condition     removes  the unphysical modes corresponding to gauge field
  from the total Fock space. The Gupta-Bleuler
type condition is used to remove the unphysical gaugeon modes from the physical states. 
The second subsidiary condition is valid, when
  $Y_\nu^\star$   satisfies
the following free equation:
\begin{eqnarray}
(\partial_\mu \partial^\mu  -m^2)Y_\nu^\star  =0,\label{fr}
\end{eqnarray}
which we have derived using equations of motion.
The  free equation  (\ref{fr}) guarantees the decomposition of  $Y_\nu^\star$ 
into 
positive and negative frequency parts. Consequently, the subsidiary conditions (\ref{con}) 
warrant the positivity of the semi-definite
 metric of our physical state-vector space:
\begin{equation}
\langle \mbox{phys}| \mbox{phys}\rangle\geq 0,
\end{equation} 
and hence, we have a desirable physical subspace  on which our unitary physical
$S$-matrix exists.

\subsection{BRST symmetric gaugeon formalism }
In this subsection we discuss the BRST symmetric gaugeon formalism for
Abelian 2-form gauge theory, with the  Lagrangian density:
 \begin{eqnarray}
S_{BY} &=&\int d^dx \left[ \frac{1}{12}  \tilde F_{\mu \nu \rho}\tilde F^{\mu \nu \rho}  -i\tilde \partial_\mu\bar\rho_\nu (\tilde \partial^\mu\rho^\nu -
\tilde \partial^\nu\rho^\mu )+\tilde \partial_\mu\bar\sigma\tilde \partial^\mu\sigma +\beta_\nu(\tilde \partial_\mu B^{
\mu\nu}   -\tilde \partial^\nu\varphi)\right.\nonumber\\ 
&+&\left. \epsilon(Y^\star_\nu +\alpha\beta_\nu)^2 -(\tilde \partial_\mu Y^\star_\nu -
\tilde \partial_\nu Y^\star_\mu )\tilde \partial^\mu Y^\nu - i\bar\chi\tilde \partial_\mu\rho^\mu -i\chi (\tilde \partial_\mu\bar\rho^\mu -
\lambda_2\bar\chi)\right.\nonumber\\ 
&-& \left. i\tilde \partial_\mu K^\star_\nu (\tilde \partial^\mu K^\nu -
\tilde \partial^\nu K^\mu )+ \tilde \partial_\mu Z^\star \tilde \partial^\mu Z\right]\nonumber\\
&=&\int d^dx \left[\frac{1}{12}  \tilde F_{\mu \nu \rho}\tilde F^{\mu \nu \rho}    -
i\bar\rho_\nu \left(\partial_\mu\partial^\mu \rho^\nu -\partial_\mu\partial^\nu
\rho^\mu -m^2\rho^\nu +\frac{1}{2}	\frac{m^2}{n\cdot \partial}n^\nu\partial\cdot\rho
+ \frac{1}{2}	\frac{m^2}{n\cdot \partial} \partial^\nu n\cdot\rho\right.\right.\nonumber\\
& -&\left.\left. \frac{1}{4}\frac{m^4}{(n\cdot\partial)^2}n^\nu n\cdot\rho\right) -\bar{\sigma}
(\partial_\mu\partial^\mu -m^2)\sigma +\beta_\nu\partial_\mu B^{
\mu\nu} -\frac{1}{2}m^2\beta_\nu\frac{1}{n\cdot\partial}n_\mu B^{\mu\nu}+\epsilon(Y^\star_\nu +\alpha\beta_\nu)^2 \right.\nonumber\\
&-& \left. \beta_\nu\partial^\nu\varphi +  \frac{1}{2}	\frac{m^2}{n\cdot \partial} \beta_\nu
  n^\nu\varphi  -  (\partial_\mu Y^\star_\nu - \partial_\nu Y^\star_\mu )   \partial^\mu Y^\nu +\frac{1}{2}  (\partial_\mu Y^\star_\nu - \partial_\nu Y^\star_\mu )\frac{m^2}{n\cdot \partial} n^\mu Y^\nu   \right.\nonumber\\
&+& \left. \frac{1}{2}\frac{m^2}{n\cdot \partial} (n_\mu Y_\nu^\star -n_\nu Y_\mu^\star) \partial^\mu Y^\nu  -\frac{1}{4}\frac{m^2}{n\cdot \partial} (n_\mu Y_\nu^\star -n_\nu Y_\mu^\star)  \frac{m^2}{n\cdot \partial} n^\mu Y^\nu  -i\bar\chi \partial_\mu\rho^\mu +\frac{i}{2}m^2\bar\chi\frac{1}{n\cdot\partial}n_\mu \rho^\mu\right.\nonumber\\
&-& \left.  i\lambda_2\bar\chi\chi -i\bar\rho^\mu\partial_\mu\chi-\frac{i}{2}\frac{m^2}{n\cdot\partial}\bar\rho^\mu n_\mu\chi -i\partial_\mu K_\nu^\star (\partial^\mu K^\nu -\partial^\nu K^\mu) +\frac{i}{2} \partial_\mu K_\nu^\star\frac{m^2}{n\cdot \partial}(n^\mu K^\nu -\partial^\nu K^\mu )  \right.\nonumber\\
&+& \left.\frac{i}{2}\frac{m^2}{n\cdot\partial } n_\mu K_\nu^\star
(\partial^\mu K^\nu -\partial^\nu K^\mu)+ \frac{i}{4}\frac{m^2}{n\cdot\partial }n_\mu K_\nu^\star\frac{m^2}{n\cdot\partial } n^\nu K^\mu - Z^\star (\square -m^2)Z \right],\label{yh} 
\end{eqnarray}
 where $K_\nu, K^\star_\nu$ and $Z, Z^\star$ are the ghost fields and ghost of ghost fields, corresponding to the 
 gaugeon fields.

The gaugeon fields and respective ghost and ghost of ghost fields change  under the  BRST transformations:
\begin{eqnarray}
s_b Y_\nu &=&K_\nu,\ \ \
s_b K_\nu =0,\nonumber\\
s_b K_\nu^\star &=& iY_\nu^\star,\ \
s_b Y_\nu^\star =0,\nonumber\\
s_b Z^\star &=&0,\ \ s_b Z=0.\label{brs}
\end{eqnarray}
Therefore, the gaugeon Lagrangian density (\ref{yh}) remains intact under
 the effect of combined  BRST transformations  (\ref{brst11}) and (\ref{brs}).
 
 The BRST charge is given by
 \begin{eqnarray}
Q_B &=&\int d^{d-1}x \left[-2\tilde F^{0\nu \rho} (\tilde\partial_0\rho_\nu - \tilde\partial_\nu\rho_0) + \beta_\nu (\tilde\partial^0\rho^\nu - \tilde\partial^\nu\rho^0) - \tilde\partial_\nu \sigma (\tilde\partial^0\bar\rho^\nu - \tilde\partial^\nu\bar\rho^0)\right. \nonumber\\ 
&+& \left.\bar\chi \tilde\partial^0\sigma - \chi B^0 - K_\nu(\tilde\partial^0 Y^{\star\nu} - \tilde\partial^\nu Y^{\star 0}) + Y^\star_\nu(\tilde\partial^0 K^\nu - \tilde\partial^\nu K^0)\right],
\end{eqnarray}
 which  annihilates the physical subspace of
 of the total Hilbert space:
\begin{eqnarray}
Q_B|\mbox{phys}\rangle =0.
\end{eqnarray}
 This single subsidiary condition of Kugo-Ojima type removes both  the
 unphysical gauge modes as well as unphysical gaugeon modes. 
 
The gaugeon Lagrangian density  (\ref{yh}) also admits the following quantum gauge transformations:
\begin{eqnarray}
 &&\delta_q  B_{\mu\nu}  = \tau\left(  \partial_\mu Y_\nu -  \partial_\nu Y_\mu -\frac{1}{2}\frac{m^2}{n\cdot\partial}n_\mu Y_\nu +\frac{1}{2}\frac{m^2}{n\cdot\partial}n_\nu Y_\mu\right),  \nonumber\\
&&\delta_q  \rho_\mu =\tau K_\nu,\ \
 \delta_q  \sigma =\tau Z, \nonumber\\
&&\delta_q  Y_\nu^\star = -\tau \beta_\nu,  \ \
   \delta_q  K_\mu^\star = -\tau \bar\rho_\mu,  \nonumber\\
   &&\delta_q  Z^\star = -\tau \bar\sigma,\ \
    \delta_q \Theta  =0,  \nonumber\\ &&\Theta = \bar\rho_\mu, \beta_\mu, \bar\sigma, \bar\chi, \varphi, \chi, Y_\nu. \label{qg}
\end{eqnarray}
 Under the above quantum gauge transformation, Lagrangian density (\ref{yh}) is 
 form-invariant:
 \begin{eqnarray}
S_{BY}(\phi^A ,\alpha) =  S_{BY}(\hat\phi^A ,\hat\alpha),
\end{eqnarray}
 where
 \begin{equation}
 \hat\alpha  =\alpha +\tau \alpha.\label{alp1}
\end{equation}
Here `` $\hat{}$ " refers to the quantum gauge transformed quantity. 
It is easy to see that, the quantum gauge transformations  in the 2-form gauge theory  (\ref{qg}) also  commute  with BRST transformations mentioned  in  (\ref{brs}). Consequently, it is confirmed that the Hilbert space, spanned from
 physical states, annihilated by BRST charge, is also invariant under the quantum gauge
transformations:
\begin{equation}
\hat Q_B = Q_B.
\end{equation}
 Hence, the physical subspace in case of Abelian 2-form gauge theory is also invariant under quantum gauge transformation.
 \section{Conclusion}
 In conclusion, we have investigated the gaugeon formalism in the context of VSR.
It is  found that the gaugeon modes together with gauge modes become massive in the VSR scenario. Our results are very general and will be valid for any (arbitrary)  gauge
theory. For illustration, we have first considered Maxwell theory in the VSR framework, 
 which remains invariant under the  modified gauge transformations, rather
 being invariant under usual gauge transformations. For a VSR-type Lorentz gauge, we have obtained a Proca type equation for Euler-Lagrange equation of motion, revealing 
 that the gauge field gets a mass.  Further, to investigate
 quantum gauge symmetry, we introduced the gaugeon modes to the VSR invariant Maxwell theory.  Remarkably, the subsidiary condition to remove unphysical
 gaugeon modes satisfy the Proca type equation.
 This observation confirms that the   gaugeon fields are also massive, where as
 in SR invariant theory, these  gaugeon fields were massless.
 Therefore, the massive gaugeon fields could possibly be a candidate for 
 dark matter, where mass appears naturally due to VSR effect. Further, 
 we have studied the BRST symmetric gaugeon formalism and  obtained only one (most general) subsidiary condition of Kugo-Ojima type.
Analogously,  we have derived the gaugeon formalism for Abelian 2-form gauge theory also in VSR. Here, we have found that there exist many
more auxiliary fields, as it is a reducible theory
and all the modes become massive with same mass. This is the reason why this
mass generation is different from Higgs mechanism.  It would be interesting 
to explore the present investigation for the higher-form gauge theories.
The higher-form fields are important ingredients to certain string and super-gravity theories.
From the viewpoint that the VSR symmetry is a fundamental symmetry of nature, there
is no obvious reason to stop introducing additional VSR-invariant but Lorentz-violating
interactions to  these theories.

Other approaches for mass generation have also been studied  \cite{pani,lah},  
involving the topological terms. 
However, a   massive Proca theory can be imbibed with gauge invariance   by introducing
a scalar mode. An Abelian gauge theory describing dynamics of massive spin one bosons is
studied recently, where it is found that, the theory respects
Lorentz invariance, locality, causality and unitarity \cite{viv}.
In Ref. \cite{shuk,shuk1},  the possibility to attain scale invariance and novel symmetries
for the massive theory, through suitable coupling
with scalar-tensor gravity are demonstrated. In this process, not only the 
 photon  becomes massive, but the scalar-tensor gravity also acquires non-vanishing conserved scale current. It will be fascinating to explore  such a possibility from the VSR point of view.

 The structure of results we obtained here by studying   gaugeon formalism in VSR is   not very different to that of SR case.  Unlike  the SR case, the novel observation  is that in VSR scenario,  all the fields (including gaugeon and ghost fields) acquire mass, which modifies the masses of the original dispersion relations.
 Present investigation might play an important role for  discrete symmetry violating  gauge theories.
 In a very recent work, following Sakoda's treatment of Yang-Mills fields \cite{sako},  
the Harada-Tsutsui gauge recovery procedure \cite{hara} has been applied to the gauge 
non-invariant functional and Type I  and the extended Type I gaugeon formalism
have been obtained \cite{en}. We have studied the Sakoda's treatment for
1-form gauge theory in VSR by  including the two gauges of the standard formalism. 
In this consideration, the  theory describing extended gauge freedom
occupies the total Fock space, which embeds the
Fock spaces of both gauges. It has been found that  the Harada-Tsutsui gauge recovery procedure still holds for VSR scenario. We have found that Sakoda's effective action
coincides with the gaugeon action.   However, 
  Sakoda's theory cannot arbitrarily change the gauge parameter 
  as is done by gaugeon formalism.

  From a quantum field theoretic structure suitable to describe the dark matter, it has been 
  emphasized,  in Ref. \cite{dv}, that    VSR plays  the same  role for the dark matter 
 fields as  SR does for the standard model fields. From this perspective,
our present investigations will also play an important role  in clear  understanding of the dark matter, where 
massive gaugeon and gauge fields as the dark matter candidates support  the algebraic structure underlying VSR.
 Possible physical realization of these novel form of mass generation is also of interest.

\end{document}